\documentclass[11pt]{article}

\usepackage{amsmath,amssymb}

\newcommand{\Td}[2]{\frac{d{#1}}{d{#2}}}
\newcommand{\Pd}[2]{\frac{\partial{#1}}{\partial{#2}}}

\newcommand{\ltd}[1]{\frac{d}{d{#1}}}

\newcommand{\del}{\partial}

\newcommand{\half}{\frac{1}{2}}

\newcommand{\kahler}{K\"{a}hler }

\newcommand{\bsubeq}{\begin{subequations}}
\newcommand{\esubeq}{\end{subequations}}
\newcommand{\vs}[1]{\vspace{#1 mm}}

\newcommand {\beq}{\begin{eqnarray}}
\newcommand {\eeq}{\end{eqnarray}}

\newcommand {\1}[1]{\frac{1}{#1}}

\newcommand {\ph}{\varphi}

\newcommand {\sig}{\sigma}

\begin{document}

\allowdisplaybreaks{

\setcounter{page}{0}

\begin{titlepage}

{\normalsize
\begin{flushright}
OU-HET 390\\
{\tt hep-th/0107100}\\
July 2001
\end{flushright}
}
\bigskip

\begin{center}
{\LARGE\bf A Note on Conifolds}

\vs{10}

\bigskip
{\renewcommand{\thefootnote}{\fnsymbol{footnote}}
{\Large\bf Kiyoshi Higashijima\footnote{
     E-mail: {\tt higashij@phys.sci.osaka-u.ac.jp}},
 Tetsuji Kimura\footnote{
     E-mail: {\tt t-kimura@het.phys.sci.osaka-u.ac.jp}} {\large and}
 Muneto Nitta\footnote{
     E-mail: {\tt nitta@het.phys.sci.osaka-u.ac.jp}}
}}

\setcounter{footnote}{0}
\bigskip

{\large\sl
Department of Physics,
Graduate School of Science, Osaka University, \\
Toyonaka, Osaka 560-0043, Japan \\
}
\end{center}
\bigskip


\begin{abstract}

We present the Ricci-flat metric and its \kahler potential 
on the conifold with the $O(N)$ isometry, 
whose conical singularity is repaired by 
the complex quadric surface 
$Q^{N-2} = SO(N)/SO(N-2)\times U(1)$.

\end{abstract}

\end{titlepage}


{\sl Introduction}.  
Conformally invariant nonlinear sigma models with 
${\cal N}=2$ supersymmetry in two-dimensions 
describe the superstring in curved space. 
The target space must be a Ricci-flat \kahler manifold 
by the requirement of finiteness~\cite{AG,GVZ,NS}. 
In the previous letter~\cite{HKN1}, 
we presented the simple derivation 
of the Ricci-flat metric on the deformed conifold 
with the $O(N)$ isometry, 
whose conical singularity is removed by $S^{N-1}$. 
It coincides with the Stenzel metric on the 
cotangent bundle over $S^{N-1}$~\cite{St}, 
and includes the Eguchi-Hanson gravitational 
instanton~\cite{EH} 
and the six-dimensional deformed conifold~\cite{CO,OY} 
in the cases of $N=3$ and $N=4$, respectively.
The metric contains the deformation parameter, 
and the manifold becomes a conifold 
when the parameter vanishes.

In this letter, we present the explicit form of 
the Ricci-flat metric and its \kahler potential on 
the conifold, whose conical singularity is repaired by 
the {\it complex quadric surface}  
$Q^{N-2} \equiv SO(N)/SO(N-2) \times U(1)$.  
It contains a resolution parameter $b$ as an integration constant, 
which controls the size of $Q^{N-2}$. 
The limit of $b \to 0$ corresponds to the conifold, 
which coincides with the singular limit of 
the deformed conifold. 
Our manifold can be interpreted as 
the line bundle over $Q^{N-2}$. 
The four-dimensional manifold of $N=3$ is 
again the Eguchi-Hanson space, 
in which the conical singularity is removed by 
$Q^1 \simeq S^2$.  
In the case of the six-dimensional manifold of $N=4$, 
the conical singularity is repaired by 
$Q^2 \simeq S^2 \times S^2$, 
and it gives a way to repair the singularity 
different from the {\it deformation} by $S^3$~\cite{CO,OY} or 
the so-called {\it small resolution} by $S^2$~\cite{CO,PT}.

\medskip
{\sl Definition of the model}. 
${\cal N}=2$ supersymmetric nonlinear sigma models in
two dimensions are described by the chiral superfields 
$\ph^{\alpha}(x,\theta,\bar\theta)$ and 
the \kahler potential ${\cal K}(\ph,\ph^*)$~\cite{Zu}. 
The Lagrangian is given by 
${\cal L}= \int d^4\theta \, {\cal K} = 
g_{\alpha\beta^*}(\ph,\ph^*)
\del_{\mu}\ph^{\alpha}\del^{\mu}\ph^{*\beta} + \cdots$, 
where the \kahler metric is defined by 
$g_{\alpha\beta^*} = \del_{\alpha}\del_{\beta^*}{\cal K}$ 
with $\del_{\alpha} = \del / \del \varphi^{\alpha}$ and
$\del_{\alpha^*} = \del / \del \varphi^{* \alpha}$.  
(Here we have used the same letters for chiral superfields and 
their components.) 

First, we prepare chiral superfields 
$\phi^A(x,\theta,\bar\theta)$ ($A = 1, 2, \cdots, N$; $N \geq 3$), 
constituting the vector $\vec{\phi}(x,\theta,\bar\theta)$ of $O(N)$. 
We define the $O(N)$ symmetric target space 
by imposing the constraint
\beq
 \sum_{A=1}^N (\phi^A)^2 \ = \ 0\,. \label{con1.}
\eeq
This constraint defines the conifold with 
the real dimension $2N-2$.
We can rewrite this by an unitary transformation as 
\beq 
 \vec{\phi}^T J \vec{\phi} \ = \ 0. \label{con.}
\eeq
Here $J$ is the rank-$2$ invariant tensor of $O(N)$, 
which we take as  
\begin{align}
 J \ &= \ \left(
 \begin{array}{ccc}
       0 & {\bf 0}       & 1 \\
 {\bf 0} & {\bf 1}_{N-2} & {\bf 0} \\
       1 & {\bf 0}       & 0
\end{array} \right) \; .
\end{align}
where ${\bf 1}_{N-2}$ is the $(N-2) \times (N-2)$ unit matrix. 

Introducing an auxiliary chiral superfield 
$\phi_0(x,\theta,\bar \theta)$, 
we can give the $O(N)$ symmetric Lagrangian by
\begin{align}
{\cal L} 
\ &= \ \int \! d^4 \theta \,{\cal K} (X) 
+ \Big( \int \! d^2 \theta \; \phi_0 \vec{\phi}^T J \vec{\phi} 
+ {\rm c.c.} \Big) \; .\label{Lagrangian}
\end{align}
Here, $X(x,\theta,\bar\theta)$ 
is the $O(N)$-invariant real superfield, defined by 
\begin{align}
 X \ &\equiv \ \sum_{A=1}^N \phi^{\dagger A} \phi^A \; , 
\end{align}
and ${\cal K}(X)$ is an arbitrary function of $X$. 
The symmetry of the Lagrangian (\ref{Lagrangian}) 
is $G = O(N) \times U(1)$, assigning 
the $U(1)$ charges of $\phi^A$ and $\phi_0$,   
$1$ and $-2$, respectively. 
By the integration over the auxiliary field $\phi_0$, 
we obtain the constraint (\ref{con.}), 
which can be immediately solved as 
\begin{align}
 \vec{\phi} \ &= \ \sigma \left(
\begin{array}{c}
 1 \\
 z^i \\
 - \half (z^i)^2 
\end{array} \right) \; , \label{f-chiral}
\end{align} 
where the summation over the repeated indices is implied. 
Here $\sigma(x,\theta,\bar\theta)$ and  
$z^i(x,\theta,\bar\theta)$ ($i = 1,2,$ $\cdots,$ $N-2$) 
are chiral superfields, with the $U(1)$ charges 
$1$ and $0$, respectively. 
Scalar components of these superfields parameterize the target space, 
and the symmetry $G$ acts on those fields as 
a holomorphic isometry. 
The invariant $X$ becomes 
\begin{align}
 X 
\ = \ |\sigma|^2 
 \left[ 1 +|z^i|^2 + \frac{1}{4} (z^i)^2 (z^{* j})^2 \right] 
 \ \equiv \ |\sigma|^2 Z \; .
\end{align}

Note that the constraint (\ref{con1.}) or (\ref{con.}) 
is invariant under the complex extension of the symmetry $G$. 
Using this, any point $\vec{\phi}$ on the manifold 
can be transformed to 
$\langle \vec{\phi}^T \rangle = (1,0,\cdots, 0)$,  
which can be interpreted as 
the vacuum expectation value. From this,  
we find the symmetry $G$ is spontaneously 
broken down to $H=O(N-2)\times U(1)$. 
Hence there appear the Nambu-Goldstone bosons, 
parameterizing $G/H \simeq SO(N)/SO(N-2)$.  
The whole target manifold can be locally regarded as 
${\bf R} \times SO(N)/SO(N-2) \simeq 
{\bf R}\times S^{N-1}\times S^{N-2}$.   

\medskip
{\sl Ricci-flat Condition and Its Solution}. 
We would like to determine the function ${\cal K} (X)$, 
imposing the Ricci-flat condition on the manifold.
We use the {\it same} letters for superfields 
{\it and} their lowest components from now on.  
The \kahler metric is 
\begin{align}
 g_{\alpha \beta^*} (\ph,\ph^*)
 \ &= \ {\del^2 {\cal K}(X) 
 \over \del\ph^{\alpha}\del \ph^{* \beta}}  
 \ = \ \frac{d^2 {\cal K}}{d X^2} \Pd{X}{\varphi^{\alpha}}
\Pd{X}{\varphi^{* \beta}} + \Td{{\cal K}}{X} 
\frac{\del^2 X}{\del \varphi^{\alpha} \del \varphi^{* \beta}} \; ,
\label{metric}
\end{align}
where $\varphi^{\alpha} \equiv ( \sigma , z^i )$. 
The Ricci form is given by 
$(Ric)_{\alpha \beta^*} = 
- \del_{\alpha} \del_{\beta^*} \log \det g_{\gamma \delta^*}$,    
and the Ricci-flat condition 
$(Ric)_{\alpha \beta^*} = 0$ implies
$\det g_{\alpha \beta^*} = (\mbox{constant}) \times |F|^2$, 
with $F$ being a holomorphic function. 
This is a partial differential equation,  
which is difficult to solve in general.
The determinant $\det g_{\alpha \beta^*}$ can be 
calculated as 
\begin{align}
\det g_{\alpha \beta^*} 
\ &= \ \frac{X}{|\sigma|^2} \Big( X \frac{d^2 {\cal K}}{d X^2} 
                + \frac{d {\cal K}}{d X} \Big) 
 \Big( |\sigma|^2 \frac{d {\cal K}}{d X} \Big)^{N-2} \cdot 
 \det ( \del_i \del_{j^*} Z - Z^{-1} \del_i Z \del_{j^*} Z) \; ,
\end{align}
where $\del_i$ denotes the differentiation with respect to $z^i$: 
$\del_i Z = z^{*i} + \half z^i (z^{*j})^2$ and $\del_i
\del_{j^*} Z = \delta_{i j} + z^i z^{*j}$.
Using the complex extension of the isotropy $H$, $SO(N-2,{\bf C})$,  
we can choose a point labeled by  
$z^1 \neq 0$ and $z^m = 0$ ($m = 2, 3, \cdots, N-2$), 
without loss of generality. 
At that point, we find
\begin{align}
\det ( \del_i \del_{j^*} Z - Z^{-1} \del_i Z \del_{j^*} Z) 
 \ = \ \det \delta_{i j} \ = \ 1 \; , \ \ \ 
 X \ = \ |\sig|^2 \left(1+ {|z^1|^2 \over 2}\right)^2 \; , 
\end{align}
and then obtain
\begin{align}
\det g_{\alpha \beta^*} 
\ = \ 
|\sigma|^{2 N-6} \Big( \frac{d {\cal K}}{d X} \Big)^{N-2} 
\Big( X^2 \frac{d^2 {\cal K}}{d X^2} + X \frac{d {\cal K}}{d X}
\Big) \; . 
\label{determinant}
\end{align}
Therefore, the Ricci-flat condition becomes 
an ordinary differential equation:
\begin{align}
 \Big( \frac{d {\cal K}}{d X} \Big)^{N-2} 
 \Big( X^2 \frac{d^2 {\cal K}}{d X^2} + X \frac{d {\cal K}}{d X}
  \Big)
 \ &= \ \frac{1}{N-1} X^2 \ltd{X} 
 \left[\Big( \Td{{\cal K}}{X} \Big)^{N-1}\right]  
 + X \Big( \Td{{\cal K}}{X} \Big)^{N-1} \ \equiv \ c \; ,
\end{align}
where $c$ is a constant.
This can be immediately solved as 
\begin{align}
 \Td{{\cal K}}{X} 
\ &= \ { (\lambda X^{N-2} + b )^{\1{N-1}}\over X}  \; ,
\label{RF-sol}
\end{align}
where $\lambda$ is a constant related to $c$ and $N$, 
and $b$ is an integration constant.  
We impose $b \geq 0$ and $\lambda > 0$ 
in order that the \kahler potential is real.

The solution (\ref{RF-sol}) is sufficient to 
obtain the Ricci-flat metric using (\ref{metric}), 
but we can calculate its \kahler potential by integrating 
(\ref{RF-sol}): 
\begin{align}
{\cal K} (X) 
 \ &= \ 
\frac{N-1}{N-2} \left[ 
 (\lambda X^{N-2} + b)^{\frac{1}{N-1}} + b^{\frac{1}{N-1}} 
\cdot 
 I \Big( b^{\1{1-N}} 
      \left(\lambda X^{N-2} + b \right)^{\frac{1}{N-1}}
         ; N -1 \Big) 
 \right] \; , \label{cfd-wo-sin}
\end{align}
where the function $I(y;n=N-1)$ is defined by
\begin{align}
I (y; n) 
\  \equiv \ 
  \int^{y} \! \frac{dt}{t^n - 1} 
\ &= \ \frac{1}{n} \Big[ \log \big( y - 1 \big) 
    - \frac{1 + (-1)^n}{2} 
    \log \big( y + 1 \big) \Big] \nonumber \\
& \ \ \ \ 
 + \frac{1}{n} \sum_{r=1}^{[\frac{n-1}{2}]} \cos \frac{2 r \pi}{n} 
\cdot \log \Big( y{}^2 - 2 y \cos \frac{2 r \pi}{n} + 1 \Big) 
\nonumber \\
\ & \ \ \ \ + \frac{2}{n} \sum_{r=1}^{[\frac{n-1}{2}]} 
\sin \frac{2 r \pi}{n} 
\cdot \arctan \Big[ \frac{\cos (2 r \pi / n) - y}
                    {\sin (2 r \pi /n) } \Big] \; .
\end{align}
If we set $b=0$ in (\ref{cfd-wo-sin}), 
it becomes the \kahler potential of the conifold, 
which coincides with the one 
of the singular limit of 
the deformed conifold~\cite{HKN1}.

\medskip
{\sl Ricci-flat Metric}.
Using Eqs.~(\ref{metric}) and (\ref{RF-sol}), 
the components of the Ricci-flat metric can be calculated, to give 
\bsubeq
\begin{align}
g_{\sigma \sigma^*} 
\ &= \ 
\lambda \Big( \frac{N-2}{N-1} \Big) 
\big(\lambda X^{N-2} + b \big)^{\frac{2-N}{N-1}} X^{N-2} 
|\sigma|^{-2} \; , \\
g_{\sigma j^*} 
\ &= \ 
\lambda \Big( \frac{N-2}{N-1} \Big) 
\big(\lambda X^{N-2} + b \big)^{\frac{2-N}{N-1}} X^{N-3} 
\sigma^* \del_{j^*} Z \; , \\
g_{i j^*} 
\ &= \ 
\lambda \Big( \frac{N-2}{N-1} \Big) 
\big(\lambda X^{N-2} + b \big)^{\frac{2-N}{N-1}} X^{N-4} 
|\sigma|^4 \del_i Z \del_{j^*} Z \nonumber \\
\ & \ \ \ \ 
+ \big(\lambda X^{N-2} + b \big)^{\frac{1}{N-1}} 
 ( Z^{-1} \del_i \del_{j^*} Z - Z^{-2} \del_i Z \del_{j^*} Z) \; . 
\end{align}
\esubeq
This \kahler metric is singular at 
the surface defined by $\sigma = 0$: 
$g_{\sig\sig^*}|_{\sig=0}=0$. 
However this is just a coordinate singularity 
of the coordinate system ($\sigma$, $z^i$). 
To find regular coordinates, 
let us perform a coordinate transformation 
\begin{align}
  \rho \ &\equiv \ \frac{\sigma^{N-2}}{N-2} \; , \label{co.tr.}
\end{align}
with $z^i$ being unchanged.  
The components of the \kahler metric in 
the new coordinates $(\rho,z^i)$ are 
\bsubeq
\begin{align}
g_{\rho \rho^*} \ &= \ \lambda \Big(
\frac{N-2}{N-1} \Big) \big( \lambda X^{N-2} + b \big)^{\frac{2-N}{N-1}}
Z^{N-2} \; , \\
g_{\rho j^*} 
\ &= \ \lambda \frac{(N-2)^2}{N-1} \big( \lambda X^{N-2}
+ b \big)^{\frac{2-N}{N-1}} \rho^* Z^{N-3} \del_{j^*} Z \; , \\
g_{i j^*} 
\ &= \ \lambda \frac{(N-2)^3}{N-1} \big(
\lambda X^{N-2} + b \big)^{\frac{2-N}{N-1}} |\rho|^2 Z^{N-4} \del_i Z
 {\del}_{j^*} Z \nonumber \\
\ & \ \ \ \ + \big( \lambda X^{N-2} + b \big)^{\frac{1}{N-1}} 
 (Z^{-1} \del_i \del_{j^*} Z - Z^{-2} \del_i Z \del_{j^*} Z) \; ,
\end{align}\label{well-defined-metric}
\esubeq 
where $X = |(N-2)\rho|^{2\over N-2} Z$.  
These are non-singular at the surface of $\rho=0$, 
corresponding to $\sig=0$,  
as long as the integration constant $b$ takes a non-zero value.  
In the limit of $b \to 0$, 
the manifold becomes the conifold and  
the metric (\ref{well-defined-metric}) becomes singular 
at $\rho = 0$. 
So we can regard this constant $b$ as 
a {\it resolution} parameter 
of the conical singularity. 
The coordinate singularity in the coordinates $(\sig,z^i)$ 
is due to the identification of (\ref{co.tr.}) 
as in the Calabi metric on 
the line bundle over ${\bf C}P^{N-1}$~\cite{Ca}. 

The metric of the $\rho =0$ surface itself ($d\rho=0$) is 
\begin{align}
g_{i j^*}(z,z^*) 
\ = \ b^{\frac{1}{N-1}} ( Z^{-1} \del_i \del_{j^*} Z - Z^{-2}
\del_i Z \del_{j^*} Z ) 
. \label{metric-submfd}
\end{align}
This define a \kahler submanifold 
whose \kahler potential is given by 
\begin{align}
 {\cal K}(z,z^*) \ &= \ 
 b^{\frac{1}{N-1}}
 \log \left[ 1 + |z^i|^2 + \frac{1}{4} (z^i)^2 (z^{* j})^2 \right] 
 \ = \ 
 b^{\frac{1}{N-1}} \log Z  \; , \label{QN}
\end{align}
which is the \kahler potential of 
the complex quadric surface 
$Q^{N-2} = SO(N)/SO(N-2) \times U(1)$~\cite{HN1}--\cite{HKNT}. 
Therefore we have found that the conical singularity is 
resolved by $Q^{N-2}$ of the radius $b^{\frac{1}{2(N-1)}}$. 
The manifold can be interpreted as the line bundle over $Q^{N-2}$. 
In fact it was proved in \cite{PP} that 
there exists a Ricci-flat \kahler metric on 
the line bundle over any Einstein manifold.

\medskip
{\sl Examples}. 
Let us give the more concrete expressions for 
the $N=3$ and $N=4$ cases. 
For the four-dimensional manifold of $N=3$, 
the \kahler potential (\ref{cfd-wo-sin}) becomes 
\bsubeq
\begin{align}
{\cal K} (X) \ &= \ 2 \sqrt{\lambda X + b} + \sqrt{b} 
 \log \Big( \frac{\sqrt{\lambda X + b} - \sqrt{b}}
            {\sqrt{\lambda X + b} + \sqrt{b}}
\Big) \; .
\end{align}
\esubeq
Defining $\varrho^4 = 4 (\lambda X + b)$ and $a^4 = 4 b$,
we find that this is the \kahler potential~\cite{GP} 
of the Eguchi-Hanson gravitational instanton~\cite{EH}:
\begin{align}
{\cal K} 
\ &= \ 
\varrho^2 + \frac{a^2}{2} 
\log \Big( \frac{\varrho^2 - a^2}{\varrho^2 + a^2} \Big) \; .
\end{align}
The singularity at the apex of the conifold 
is repaired by $Q^1 \simeq S^2$, and the isometry is
$SO(3)\times U(1) \simeq U(2)$. 

The \kahler potential (\ref{cfd-wo-sin}) in the six-dimensional 
manifold of $N=4$ is 
\begin{align}
{\cal K} (X) 
\ &= \ 
\frac{3}{2} (\lambda X^2 + b )^{1/3} + \frac{b^{1/3}}{4} 
\log \Big[ \frac{ \{ (\lambda X^2 + b)^{1/3} - b^{1/3}\}^3}{\lambda X^2} 
    \Big] \nonumber \\
& \ \ \ \ - \frac{\sqrt{3} b^{1/3}}{2} 
 \arctan \Big[
  \frac{ 2 (\lambda X^2 + b)^{1/3} + b^{1/3}}{\sqrt{3} b^{1/3}} 
 \Big] \; .
\end{align}
The metric in the coordinates ($\rho$, $z^1$, $z^2$) is 
represented as follows:
\bsubeq
\begin{align}
g_{\rho \rho^*} \ &= \ \frac{2 \lambda}{3} \frac{Z^2}{(\lambda X^2 +
b)^{2/3}} \; , \ \ \ 
g_{\rho j^*} \ = \ \frac{4 \lambda}{3} \frac{\rho^* Z \del_{j^*}
Z}{(\lambda X^2 + b)^{2/3}} \; , \\ 
g_{i j^*} \ &= \ \frac{8 \lambda}{3} \frac{|\rho|^2 \del_i Z
\del_{j^*} Z}{(\lambda X^2 + b)^{2/3}} + (\lambda X^2 + b)^{1/3} 
(Z^{-1} \del_i \del_{j^*} Z - Z^{-2} \del_i Z \del_{j^*} Z) \; .
\end{align}
\esubeq
The isometry of this manifold is 
$SO(4)\times U(1)\simeq SU(2)\times SU(2)\times U(1)$. 
The singularity at the apex of the conifold is repaired by
$Q^2 \simeq S^2 \times S^2$ (the radii of these two $S^2$ coincide). 
This way of repairing of the conical singularity 
is different from either  
the {\it deformation} by $S^3$~\cite{CO,OY}  
or the {\it small resolution} by $S^2$~\cite{CO,PT}   
known in the six-dimensional conifold.  

\medskip
{\sl Discussions}. 
We can obtain the \kahler potential (\ref{QN}) of 
$Q^{N-2} = SO(N)/ SO(N-2)\times U(1)$ directly, 
by gauging the $U(1)$ part of the isometry $G$ 
and performing the integration over gauge superfields.
(This is known as \kahler quotient~\cite{HKLR}, 
and actually hold for an arbitrary \kahler potential  
${\cal K}(X)$~\cite{HN2}.)  
Replacing the base manifold $Q^{N-2}$ 
by other compact manifolds in \cite{HN1,HN3},  
we can construct other Ricci-flat \kahler manifolds, 
whose conical singularity is repaired by 
those base manifolds~\cite{HKN2}. 
Since non-perturbative effects of $Q^N$ was investigated 
using the large-$N$ method in \cite{HKNT},  
the large-$N$ limit of the conifold is also interesting. 
The investigation of super-conformal field theories corresponding to
our manifolds is an interesting task.

After the completion of this work 
we were informed that the six-dimensional manifold 
in the $N=4$ case
is known in refs.~\cite{EN,PT2}.

\section*{Acknowledgements} 
We would like to thank Michihiro Naka, 
Kazutoshi Ohta and Takashi Yokono 
for valuable comments and discussions.
This work was supported in part by the Grant-in-Aid for Scientific
Research (\#13640283).


\end{document}